\input{aipcheck}
\documentclass[
    ,final            
  ]
  {aipproc}
\layoutstyle{6x9}

\usepackage{hyperref} 
\usepackage{amsmath,amssymb} 
\usepackage{array} 
\usepackage{multirow} 

\newcommand{\be}{\begin{equation}}
\newcommand{\ee}{\end{equation}}
\newcommand{\bea}{\begin{eqnarray}}
\newcommand{\eea}{\end{eqnarray}}
\def\sss{\scriptscriptstyle}
\def\Ls{{\sss L}}

\def\GD{\Gamma_{\sss D}}
\def\GS{\Gamma_{\sss S}}
\def\TBL{T_{B-L}}

\def\GD{\Gamma_{\sss D}}
\def\GS{\Gamma_{\sss S}}

\begin{document} 

\classification{12.60.-i, 98.80.Cq, 12.60.Jv, 11.15.Ex, 98.80.Ft }
\keywords{Left-Right, supersymmetry, grand unified theory, 
domain wall}

\title{Gauged $B-L$ unification and cosmology}
\author{Urjit A. Yajnik}{
address={Indian Institute of Technology, Bombay, Mumbai - 400076, India}
}

\begin{abstract}
We discuss some cosmological implications of low energy gauged $B-L$
symmetry with and without supersymmetry. Generic possibility of 
leptogenesis from a domain wall driven first order phase transition 
is shown to be a characteristic of such models.
\end{abstract}

\maketitle

\section{Introduction}
\label{sec:intro} 
There are several motivations to search for a grand unified theory.
The tantalising possibility of gauge coupling unification into a
simple gauge group is the most attractive. A dramatic signature of 
this unification, viz., proton decay has not been observed, though
the possibilities are still open \cite{goranthisproceeding}. 
The other motivations
arise out of the need to understand the known conserved charges
in a larger perspective. Quantization of electromagnetic charge 
is one goal. The other intriguing feature of low energy physics
is separate conservation of baryon and lepton numbers, without
any associated gauge fields.
This puzzle is partly resolved by the fact that the number $B+L$ 
is anomalous in the Standard Model (SM) and hence not really conserved.
This leaves us with the need to understand $B-L$ symmetry.

The astrophysical and atmospheric data on neutrinos show that
the total lepton number $L$ is conserved and that at least two 
of the neutrinos have mass. The smallness of these masses 
$\sim 10^{-3}-10^{-1}$eV,  relative to the electroweak  scale 
$250$GeV needs an explanation. See-saw mechanism which
provides an attractive explanation implies Majorana mass
and violation of total lepton number $L$ in perturbation theory 
at a high scale.

A very attractive possibility for unification is that the remaining
exact global symmetry $B-L$ is gauged. In this review we summarise
the situation for potential signatures of unification with gauged $B-L$. 
Thermal leptogenesis presents a very attractive link between low 
energy neutrino data and gauged $B-L$ symmetry. At present this 
link seems to have run into problems of energy scales as will be 
discussed. We make a case for a low energy $10^4-10^6$ GeV
gauged $B-L$ symmetry based on several of our joint previous 
works. It relies on leptogenesis from a first order phase transition and 
demonstrates its naturalness from the point of view of gauged $B-L$. 
Consistency of this scenario with other cosmological issues and 
possible  ways of verifying it are discussed.

\section{Cosmology with gauged \( B-L \)}
\label{sec:cosmo}
There is at least one cosmological reason why gauged 
$B-L$ would be an appealing feature of nature. This is the
observed baryon asymmetry of the Universe. At present three 
independent sources, direct observations, WMAP observations
and nucleosynthesis data place the value of this asymmetry,
expressed as ratio of net baryon number density to entropy density
in photons, at  $\eta_B \equiv n_B/s$ $= 6\times10^{-10}$.
Since $B+L$ is anomalous,
its equilibrium value would be zero in the early Universe above the
electroweak temperature scale $\sim 100$GeV. Thus
baryon asymmetry must be generated out of $B-L$ asymmetry. If the 
latter is only a global charge, the observed baryon asymmetry 
could have arisen out of accidental initial conditions present
at the Planck scale. A more appealing possibility is that the
$B-L$ is a gauged abelian charge. This could be intrinsically so
or arising from breaking of a more fundamental compact non-abelian
symmetry group. But an abelian charge will have natural value
zero in the early Universe. 
Any model of breaking this gauge invariance can then give masses
to heavy neutrinos and also permit a dynamical computation of the
baryon asymmetry.

A traditional scenario envisages generation of  $L$ asymmetry from 
out-of-equilibrium decay of the heavy majorana fermions \cite{fy86}. 
This asymmetry is then converted to $B$ asymmetry by electroweak
sphaleronic processes \cite{krs85} whose rate at a high temperature 
$T$ is given by
\be
\Gamma_{\mbox{sphaleron}} = A \alpha_W^5 T^4
\label{eq:sphalrate}
\ee
with $A$ a numerical constant of order unity and $\alpha_W$ is the weak
interaction coupling constant $g^2/4\pi$.
The status of thermal Leptogenesis proposal may be summarised as 
follows. The net baryon asymmetry is given by
\be
Y_B =0.55\epsilon Y_{N_1} d,
\label{eq:bau_def}
\ee
where $\epsilon$ is the $CP$ violation parameter, $Y_{N_1}$ is the
abundance of the lightest of the heavy neutrino species $N_1$ at 
the $L$-violating scale and $d$ is subsequent dilution.
Using see-saw mechanism, the Low energy neutrino data constrain 
$\epsilon$ to  remain smaller than \cite{di02}
\be 
|\epsilon| \le  9.86 \times 10^{-8}\left( \frac{M_1}{10^9 GeV} \right)
\left( \frac{m_{(3)}}{0.05eV} \right)
\ee
where $m_{(3)}$ is the mass of the heaviest of the light neutrinos.
Combining with the observed bound on Baryon asymmetry from 
WMAP gives a lower bound on $M_1$ as \cite{bbp02_a,bbp02_b}
\be 
M_1\geq O(10^9) GeV \left(\frac{2.5\times 10^{-3}}{Y_{N_1}d }
\right)\left(\frac{0.05 eV}{m_{(3)}}\right)\,.
\ee
Since the last two quantities in brackets are expected to be order 
unity, this constrains $M_1$ to be bigger than $10^9$GeV.

The high scale means direct verification of 
this symmetry is impossible in forseeable future. But more seriously,
the bound does not agree with supersymmetry (SUSY) as a potential
solution of the fundamental Higgs and hierarchy problems. Most 
supersymmetric models have a problem of overabundance of gravitinos 
 if subsequent to any scale higher than $10^9$GeV, the Universe 
has had purely radiation and matter dominated epochs. 

\section{Model independent limit on heavy neutrino mass}
\label{sec:scales}
The above constraint on the scale of heavy neutrino masses is
too stringent due to the assumption of purely thermal leptogenesis.
Leptogenesis from a first order phase transition  
provides a constraint which is significantly weaker. The only 
requirement is that the raw lepton asymmetry created by the 
first order phase transition should be erased only partially
by the majorana neutrinos. The requisite baryon asymmetry is 
generated out of the remainder by the sphaleronic effects.

The dilution factor $d$ introduced in eq. (\ref{eq:bau_def}) is 
determined by two kinds of processes. They are (i) scattering processes (S) 
among the SM fermions and (ii) Decay (D) and inverse decays (ID) 
of the heavy neutrinos. The dominant contributions 
to the two types of  processes are governed by the temperature 
dependent rates
\be
\label{eq:rates}
\GD \sim \frac{h^2 M_1^2}{16\pi(4T^2 + M_1^2 )^{1/2}}
\hspace{1cm} {\rm and} \hspace{1cm}
\GS \sim \frac{h^4}{13\pi^3}\frac{T^3}{(9T^2 + M_1^2)},
\ee
where $h$ is typical Dirac Yukawa coupling of the neutrino.

We shall see in later discussion
that the raw $B-L$ asymmetry 
generated can be $\sim O(1)$. Accordingly we parameterise the dilution 
factor in the exponential form  $d=10^{-d_{\sss B}}$, so that 
natural value of $d_{\sss B}$ ranges from  $0$ to $10$.  
The integrated dilution factors 
resulting from soution of Boltzmann equations 
can be  transformed into an upper limit on the 
light neutrino masses using the canonical seesaw relation.
First consider the case $M_1>\TBL$, where $\TBL$ is the 
temperature of the $B-L$ symmetry breaking phase transition.
In this case the dilution processes are rather ineffective and the 
condition  is that in fact  $d_{\sss B}$ should be vanishingly small. 
In this case the raw lepton  asymmetry should be produced in the required 
range of values, $\sim 10^{-9}$. $M_1$ remains unconstrained.

In the opposite regime $M_1<\TBL$, both of the above types of 
processes could freely occur. The condition that complete erasure 
is prevented requires that the above processes are slower than 
the expansion scale of the Universe for all $T>M_1$. It turns out 
to be sufficient~\cite{fglp91} to require $\GD<H$ which also 
ensures that $\GS<H$. This leads to a requirement on the
lightest neutrino mass $m_{(1)}$,
\be
m_{(1)} < m_* \equiv 4\pi g_*^{1/2}\frac{G_N^{1/2}}{\sqrt{2}G_F}
= 6.5\times10^{-4}eV
\label{FGLPbound}
\ee
where the parameter $m_*$~\cite{fglp91} contains only universal
couplings and $g_*$, the  effective number of thermodynamic degrees 
of freedom, and may be called the \emph{cosmological neutrino 
mass}. 

In a specific texture model of Dirac mass matrix $m_D$
\cite{Fritzsch:1979zq,Sahu:2004ny}
the see-saw relation reads, 
\be
m_{(1)} \simeq (2 \times 10^{-2})^2 \left(\frac{10^8 GeV}{M_1}\right) eV.
\ee
Then eq. (\ref{FGLPbound}) seems to be naturally satisfied
for $M_1\gtrsim 10^8 GeV$. However the front factor arises from the
squared Dirac mass of the charged leptons. If we assume the Dirac mass
for neutrinos be $10^{-2}$ smaller, this factor becomes $(2 \times 10^{-4})^2$.
This results in the modest bound $M_1> 10^4$GeV, on the mass 
of the lightest  of the  heavy majorana neutrinos. 

\section{Supersymmetric Left-Right symmetric models}
\label{sec:L-Rmodel}
While elegant, seesaw mechanism predicts a new high scale which gives rise 
to a hierarchy. Inclusion of 
SUSY improves the situation, stabilizing hierarchies of 
mass scales that lie above the SUSY breaking scale. We assume the
most optimistic value for the SUSY breaking scale, being the TeV
scale without disturbing the SM. We explore the possibility for SM
to have descended from a Left-Right symmetric theory, with a low scale, 
a few orders of magnitude  removed from the TeV scale.  At a higher 
energy scale the model may turn out to be embedded in  the 
supersymmetric $SO(10)$.

We have considered two possibilities for a minimal supersymmetric 
Left-Right symmetric model with the gauge group 
$SU(3)_c \otimes SU(2)_L \otimes SU(2)_R \otimes U(1)_{B-L}$. 
Both need quark and lepton superfields, one set for each generation.  
The minimal set of Higgs  superfields required is bidoublets $\Phi_1$, $\Phi_2$ 
and triplets $\Delta$, $\bar{\Delta}$, $\Delta_c$, $\bar{\Delta}_c$ as detailed
in \cite{abs97} and \cite{sy07}.
Under discrete parity symmetry the fields have two possible transformation
rules,
\begin{eqnarray}
Q \leftrightarrow Q_c^*, \quad & 
L \leftrightarrow L_c^*, \quad & 
\Phi_i \leftrightarrow \Phi^\dagger_i,  \nonumber \\
\Delta \leftrightarrow \Delta_c^*,  \quad & 
\bar{\Delta} \leftrightarrow \bar{\Delta}_c^*, \quad & 
\Omega \leftrightarrow \pm \Omega_c^*.
\label{eq:parity} 
\end{eqnarray}
We refer to the original model with $+$ sign as MSLRM. In this case
there is a need for Gauge Mediated SUSY Breaking or gravity 
induced soft terms to lift the symmetry between Left and Right vacua.
The alternative with the $-$ sign is along the lines of the non-supersymmetric
model of Chang et al. \cite{cmp84}. Here spontaneous breaking of parity 
is implemented within the Higgs structure of the theory.
We dub this model MSLR$\not\text{P}$.

In both these models, 
$SU(2)_R$ first breaks to its subgroup $U(1)_R$, at a scale $M_R$, 
without affecting the $U(1)_{B-L}$. At a lower scale $M_{B-L}$,
$SU(3)_c \otimes SU(2)_L \otimes U(1)_R \otimes U(1)_{B-L}$ breaks to
$SU(3)_c \otimes SU(2)_L \otimes U(1)_Y$. Consistency of the scheme
requires a "see-saw" relation between these two scales and the electroweak scale
$M_W$,
\be
M_{B-L}^2 \simeq M_R M_W
\ee
In this scheme parity 
is spontaneously broken while preserving electromagnetic charge invariance.
However, due to parity invariance of the original theory, the 
phenomenologically unacceptable phase 
$SU(3)_c \otimes SU(2)_R \otimes U(1)_Y'$ is quasi-degenerate with  
$SU(3)_c \otimes SU(2)_L \otimes U(1)_Y$. This quasi-degeneracy
gives rise to domain walls (DW) at the scale $M_R$, resulting in 
a first order phase transition.
 
\section{Uses of Domain Walls}
\label{sec:DWuses}
It has been shown \cite{mat00, kt05, ys07} 
that if DW dominate the evolution  of the universe  for a limited duration, 
the associated \textsl{secondary} inflation removes the gravitino and other 
dangerous relic fields like moduli typically regenerated after the primordial 
inflation \cite{ekn84}. It was shown in \cite{ys07, sy07} that this can be
successfully implemented in above class of Left-Right models in the course 
of the DW dominated first order phase transition.

Turning to leptogenesis, at least two of the Higgs expectation values in L-R model
are generically complex, thus providing natural $CP$
violation permitting all parameters in the Higgs 
potential to be real. Within the thickness  of the domain 
wall the $CP$ violating phase becomes position
dependent. Under these circumstances a formalism exists
\cite{jpt95,ck00, cjk98}, wherein the chemical potential
$\mu_\Ls$ created for the Lepton number can be computed as 
a solution of the diffusion equation whose source term
contains derivatives of the position dependent complex 
Dirac mass. In \cite{cynr02} the existence of such a position 
dependent phase was established on general grounds and verified 
in numerical simulations. The resulting value of the raw 
Lepton  asymmetry to the entropy density ratio was shown to be
\be
\eta^{\rm raw}_{\sss L} \cong 0.01\,  v_w \frac{1}{ g_*}\, 
	\frac{M_1^4}{ T^5\Delta_w}
\label{eq:ans2}
\ee
with  $\Delta_w$ standing 
for the wall width, and $v_w$ for the average wall velocity. 
A possible verification of DW based phase transition scenario
can be sought in the upcoming space based gravitational wave detectors 
capable of detecting the stochastic background arising at such phase
transitions \cite{gs07}.

An important constraint on these scenarios is that the DW must be 
metastable, with a decay temperature $T_D$  which must be larger 
than $\sim 10$ MeV in order to not interfere with Big  Bang Nucleosynthesis 
(BBN). It has been observed in \cite{ptww91} that the free energy density 
difference $\delta \rho$ between the vacua, which determines the 
pressure difference across a domain wall should be of the order 
\begin{equation}
\delta \rho \sim T_D^4
\label{eq:dr_Td_rel}
\end{equation}
in order for the DW structure to disappear at the scale $T_D$. 
We use this constraint to determine the differences between 
the relevant soft parameters
for a range of permissible values of $T_D$. The soft terms in the 
superpotential for the two models considered here have the form
\begin{eqnarray} 
\mathcal{L}_{soft} &=& 
\alpha_1 \textrm{Tr} (\Delta \Omega \Delta^{\dagger}) -
\alpha_2 \textrm{Tr} (\bar{\Delta} \Omega \bar{\Delta}^{\dagger}) +
\alpha_3 \textrm{Tr} (\Delta_c \Omega_c \Delta^{\dagger}_c) + 
\alpha_4 \textrm{Tr} (\bar{\Delta}_c \Omega_c \bar{\Delta}^{\dagger}_c) ~~~~~
\label{eq:sigNdel} \\ 
&& + ~m_1^2 \textrm{Tr} (\Delta \Delta^{\dagger}) +
m_2^2 \textrm{Tr} (\bar{\Delta} \bar{\Delta}^{\dagger}) + 
m_3^2 \textrm{Tr} (\Delta_c \Delta^{\dagger}_c) +
m_4^2 \textrm{Tr} (\bar{\Delta}_c \bar{\Delta}^{\dagger}_c) 
\label{eq:delta} \\
&& + ~\beta_1 \textrm{Tr} (\Omega \Omega^{\dagger}) +
\beta_2 \textrm{Tr} (\Omega_c \Omega^{\dagger}_c) ~.
\label{eq:omega} 
\end{eqnarray} 
In the case of MSLRM
the $\alpha$ terms do not contribute and the other terms are constrained
as listed in the table 
where we have considered 
$m_1^2 \backsimeq m_2^2 \equiv m^2$, $m_3^2 \backsimeq m_4^2 \equiv 
m^{\prime~2}$, $\langle \Omega\rangle \simeq M_R$ $\sim 10^6 $ GeV, 
$\langle\Delta\rangle \simeq M_{B-L}$ $ \sim 10^4 $ GeV, and $T_D$ in the range 
$100 \textrm{ MeV} - 10 \textrm{ GeV}$ \cite{kt05}.
%
\begin{center} 
{
\setlength\extrarowheight{1.5mm}
\begin{tabular}{c|c|c|c} 
\hline
& $T_D = 10$ GeV & $T_D = 10^2$ GeV & $T_D = 10^3$ GeV \\ \hline \hline
$(m^2 - m^{2\prime}) \sim$ & 
$10^{-4}\textrm{ GeV}^2$ & $1 \textrm{ GeV}^2$ & $10^{4} 
\textrm{ GeV}^2$\\ [1mm]
$(\beta_1 - \beta_2) \sim$ & 
$10^{-8}\textrm{ GeV}^2$ & $10^{-4}\textrm{ GeV}^2$ & 
$1 \textrm{ GeV}^2$ \\ 
[1mm] 
\hline 
\end{tabular} 
\label{tab:DWalls}}
\end{center}
The MSLR$\not\text{P}$ is more restrictive, with  $\alpha_3=-\alpha_1$,
$\alpha_4=-\alpha_2$, $m_3^2=m_1^2$, $m_4^2=m_2^2$, $\beta_1=\beta_2$.
Here only the $\alpha$ terms contribute, and for a range of temperatures 
$T_D \sim 10^2 \text{ GeV} - 10^4 \text{ GeV}$, the constraint reads 
\begin{equation} 
(\alpha_1 + \alpha_2) \sim 10^{-6} - 10^{2} \text{ GeV}.
\label{eq:a_diff} 
\end{equation} 

\section{Conclusion}
A new gauge force of nature, $U(1)_{B-L}$ may well be accessible 
in low energy accelerator data and in observable cosmological 
effects. The requirement  of thermal leptogenesis would put its 
energy scale too high to be verified. However generic occurrence 
in such models of a first order phase transition driven by domain 
walls makes low energy baryogenesis possible. If the hierarchy 
with other energy scales is protected by supersymmetry, the same 
domain walls allow for a secondary inflation to sweep  away unwanted 
relics. Supersymmetry breaking  soft parameters of such a 
model can be constrained from cosmological considerations.

\section{Acknowledgments}
\label{sec:acknowledge} 
It is a pleasure to acknowledge the collaborators with whom the
work reported here was done : J. M. Cline, S. N. Nayak, M. Rabikumar, 
Narendra Sahu and Anjishnu Sarkar. This work is supported by a 
grant from the Department of Science and Technology, India.



\end{document}